\documentclass[paper]{ieice}
\usepackage{graphicx}
\usepackage{latexsym}
\usepackage[fleqn]{amsmath}
\usepackage[varg]{txfonts}
\usepackage{subfigure}

\newcommand{\AmSLaTeX}{%
 $\mathcal A$\lower.4ex\hbox{$\!\mathcal M\!$}$\mathcal S$-\LaTeX}
\def\BibTeX{{\rmfamily B\kern-.05em
 \textsc{i\kern-.025em b}\kern-.08em
  T\kern-.1667em\lower.7ex\hbox{E}\kern-.125emX}}
\makeatletter
\def\tmpcite#1{\@ifundefined{b@#1}{\textbf{?}}{\csname b@#1\endcsname}}%
\makeatother

\field{A}
\vol{94}
\no{1}
\title[ ]
      {Outage Analysis of Cooperative Transmission with Energy Harvesting Relay: Time Switching vs Power Splitting}
\authorlist{%
 \authorentry{Guanyao Du}{ }{labelA}
 \authorentry{Ke Xiong}{ }{labelA}
 \authorentry{Zhengding Qiu}{ }{labelA}
}
\affiliate[labelA]{The author is with the School of Computer and Information Technology, Beijing Jiaotong University, Beijing 100044, P. R. China. Correspondence should be addressed to Ke Xiong; kxiong@bjtu.edu.cn}

\begin{document}
\maketitle

\begin{abstract}
Recently, energy harvesting (EH) has emerged as a promising way to realize green communications. In this paper, we investigate the multiuser transmission network with an EH cooperative relay, where a source transmits independent information to multiple destinations with the help of an energy constrained relay. The relay can harvest energy from the radio frequency (RF) signals transmitted from the source, and it helps the multiuser transmission only by consuming the harvested energy. By adopting the time switching and the power-splitting relay receiver architectures, we firstly propose two protocols, the time-switching cooperative multiuser transmission (TSCMT) protocol and the power-splitting cooperative multiuser transmission (PSCMT) protocol, to enable the simultaneous information processing and EH at the relay for the system. To evaluate the system performance, we theoretically analyze the system outage probability for the two proposed protocols, and then derive explicit expressions for each of them, respectively. Moreover, we also discuss the effects of system configuration parameters, such as the source power and relay location on the system performance. Numerical results are provided to demonstrate the accuracy of our analytical results and reveal that compared with traditional non-cooperative scheme, our proposed protocols are green solutions to offer reliable communication and lower system outage probability without consuming additional energy. In particular, for the same transmit power at the source, the PSCMT protocol is superior to the TSCMT protocol to obtain lower system outage probability.
\end{abstract}

\section{Introduction}\label{intro}

Recently, energy harvesting (EH) has emerged as a promising approach to overcome the limited energy budget of wireless networks, especially for wireless sensor networks or other networks with fixed energy supplies \cite{EH1}-\cite{EH7}. Conventional EH techniques gather energy from surrounding natural environment, for example, solar, wind, pressure, thermoelectric effects, etc. \cite{EH1}-\cite{EH4}. However, the energy obtained from physical phenomena is not always available and not easily controlled \cite{EH5}. To this end, one promising solution is to harvest energy from the ambient radio-frequency (RF) signals \cite{EH5}-\cite{EH7}, \cite{EH_ONE_WAY}-\cite{EH_Ding_POWER_ALLO}.

Cooperative communication, a technique initially proposed to offer high capacity and reliability by exploiting spatial diversity \cite{Cooeration_1}, \cite{Cooeration_2}, \cite{Cooeration_k2} has been proved to be capable of improving the energy efficiency of networks \cite{Energy_efficient_cooper_1}-\cite{Energy_efficient_cooper_3}. More recently, efforts have been made to apply EH to cooperative wireless networks to improve the performance of energy-constrained systems. As most devices used in wireless network are surrounded by RF signals (e.g., Wi-Fi signals or cellular signals), and these RF signals can carry energy and information simultaneously. Thus much attention has been paid to EH from RF signals \cite{EH_ONE_WAY}-\cite{EH_Ding_POWER_ALLO}, which is ideal for cooperative communication networks, because the transmissions of cooperation nodes can be powered by the energy harvested from the incoming signals rather than external energy supply. Specifically, in \cite{EH_ONE_WAY}, a one-way transmission among one source-destination pair was studied via an EH cooperative relay, where the achievable throughput at the destination was derived. In \cite{EH_AAN_ICC2014}, the author investigated the system achievable throughput and ergodic capacity of a decode-and-forward (DF) two-hop relaying network with an EH relay. In \cite{EH_TWO_Hop}, an amplify-and-forward (AF) two-hop transmission with the help of an EH relay was considered, where the maximal achievable information of the system were analyzed. In \cite{EH_Du}, the outage performance analysis and optimization were investigated for a DF two-way relay network with an EH relay. In \cite{EH_MULTI_COOPERA}, a cooperative uplink transmission among two users with downlink energy transfer was considered, where the system outage performance was studied. In \cite{EH_Ding}, the outage probability was characterized for users in a cooperative network where multiple source-destination pairs communicated with each other via an EH relay. In \cite{EH_Ding_POWER_ALLO}, different power allocation strategies were proposed and evaluated for the system where multiple source-destination pairs communicated with the help of a common EH relay. However, to the best of our knowledge, there has been no work investigating the multiuser transmission via an EH cooperative relay.

In this paper, we focus on the multiuser transmission network, where a source transmits independent information to multiple destinations with the help of an energy-constrained relay. The multiuser transmission network is a universal model. For example, in cellular networks, several mobile users download files from a common base station simultaneously. Moreover, if the multiuser transmission could be assisted by a wireless cooperative relay, the reliability of system can be greatly improved \cite{Multi_cooera_1}, \cite{Multi_cooera_2}. In particular, in \cite{Multi_cooera_1}, a network coding-aware cooperative relaying scheme was presented for downlink cellular networks, where two relay nodes were used to assist the transmissions for two users. In \cite{Multi_cooera_2}, an opportunistic network coding relaying cooperative scheme was proposed for a cellular downlink transmission network, where the outage performance was analyzed.\label{model}

However, all the above work did not consider EH at the relay, that is to say, the cooperative relay has to consume its own energy to assist the multiuser transmission, whereas sometimes the relay is unwilling to help due to the selfish nature or the lack of energy supply. In our work, we also focus on the multiuser transmission network, where a cooperative relay is applied to assist the multiuser transmission. Compared with previous work \cite{EH_Ding}, \cite{Multi_cooera_1}, \cite{Multi_cooera_2}, some differences of our work are deserved to be stressed as follows. Firstly, we apply EH to the cooperative relay. The EH relay can harvest energy from the RF signals it received from the source, and uses all the harvested energy to cooperate the information transfer. Secondly, in \cite{EH_Ding}, the authors considered a cooperative network with multiple source-destination pairs communicating with each other via an EH DF relay, and the impact of spatial randomness of user locations on the system outage probability was studied, whereas in this work, we aim to investigate the performance gain that the EH AF relay brings compared with the traditional non-cooperative transmission, and focus on the effect of relay position on the system outage probability.

Our main contributions can be summarized as follows. Firstly, two transmission protocols, i.e., time switching-based cooperative multiuser transmission (TSCMT) protocol and power splitting-based cooperative multiuser transmission (PSCMT) protocol, are proposed by applying the practically realizable receiver architectures in \cite{EH7} to enable the simultaneous information processing and EH at the AF relay. Secondly, for each proposed protocol, we theoretically analyze the system outage performance and derive an explicit expression for the system outage probability. As the outage probability is one of the most important performance metrics for the cooperative networks, there have been lots of works investigating the outage performance in cooperative systems \cite{EH_Ding}, \cite{EH_Ding_POWER_ALLO}, \cite{xiong_globe2013_AF}-\cite{Fan_JCS}. Thirdly, based on the analytical outage probability, we discuss the effects of system configuration parameters, such as the source power and relay location on the system performance. Extensive numerical results show that the two proposed protocols outperform the traditional non-cooperative scheme in term of outage probability. Moreover, due to the fact that the energy used by the relay is harvested from the RF signals in communication networks, our proposed protocols can improve the system outage performance without consuming extra energy.

The rest of the paper is organized as follows. Section~\ref{model} describes the system model. Section~\ref{TSCMT} and Section~\ref{PSCMT} present the proposed TSCMT and PSCMT protocols, and analyze the system outage performance for each protocol, respectively. In Section~\ref{simulation}, we provide numerical results. Finally, the conclusion is followed in Section~\ref{conclusion}.

\section{System Model}
\begin{figure}
\centering
\includegraphics[width=0.35\textwidth]{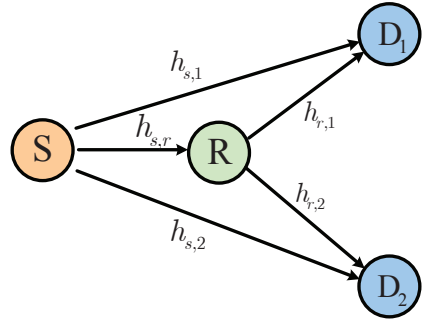}
\caption{System model and parameters.}
\label{figsystemmodel}
\end{figure}

Consider a multiuser cooperative transmission network composed of a source ${\rm{S}}$, two destinations (referred to as ${\rm{D_1}}$ and ${\rm{D_2}}$) and an energy- constrained relay  ${\rm{R}}$, as shown in Fig.~\ref{figsystemmodel} We assume that ${\rm{S}}$ has its own internal energy source and wants to transfer independent information $x_1$ and  $x_2$ to ${\rm{D_1}}$ and ${\rm{D_2}}$ with the help of ${\rm{R}}$, respectively. The energy constrained ${\rm{R}}$ relies on external charging, thus it harvests energy from the received RF signals transmitted from ${\rm{S}}$, and use all the harvested energy to help the transmissions from ${\rm{S}}$ to ${\rm{D_1}}$ and ${\rm{D_2}}$. We also assume that all the terminals have a single antenna and operate in a half-duplex mode.

Let $h_{s,r}$, $h_{s,1}$, $h_{s,2}$, $h_{r,1}$ and  $h_{r,2}$ denote the complex channel coefficient of ${\rm{S}}$ to ${\rm{R}}$ channel, ${\rm{S}}$ to ${\rm{D_1}}$ channel, ${\rm{S}}$ to ${\rm{D_2}}$ channel, ${\rm{R}}$ to ${\rm{D_1}}$ channel and ${\rm{R}}$ to ${\rm{D_2}}$ channel respectively. We assume that all the channels are quasi-block fading channel, following Rayleigh fading. Also, the channels are modeled as follows: $h_{s,r} \sim \mathcal{CN}(0,\Omega _{s,r})$, $h_{s,1} \sim \mathcal{CN}(0,\Omega _{s,1})$, $h_{s,2} \sim \mathcal{CN}(0,\Omega _{s,2})$, $h_{r,1} \sim \mathcal{CN}(0,\Omega _{r,1})$, and  $h_{r,2} \sim \mathcal{CN}(0,\Omega _{r,2})$. Specifically, let $d_{s,r}$, $d_{s,1}$, $d_{s,2}$, $d_{r,1}$ and  $d_{r,2}$ denote the distance from ${\rm{S}}$ to ${\rm{R}}$, from ${\rm{S}}$ to ${\rm{D_1}}$, from ${\rm{S}}$ to ${\rm{D_2}}$, from ${\rm{R}}$ to ${\rm{D_1}}$, and from ${\rm{R}}$ to ${\rm{D_1}}$, respectively. As a result, ${\Omega _{s,r}} = d_{s,r}^{ - m}$, ${\Omega _{s,1}} = d_{s,1}^{ - m}$, ${\Omega _{s,2}} = d_{s,2}^{ - m}$, ${\Omega _{r,1}} = d_{r,1}^{ - m}$ and ${\Omega _{r,2}} = d_{r,2}^{ - m}$, where $m$ denotes the path loss exponent.

With such a system model and assumptions mentioned above, we will describe our proposed two cooperative protocols in Section~\ref{TSCMT} and Section~\ref{PSCMT}.
\section{Time Switching-based Cooperative Multiuser Transmission (TSCMT) Protocol}
\label{TSCMT}
In this section, we consider the time switching receiver architecture proposed in \cite{EH7}. We shall first detail the proposed cooperative protocol, and then analyze the system outage performance for it \footnote{Energy harvesting from RF signal, which is also known as simultaneous wireless information and energy transfer (SWIET) was first proposed in 2008 \cite{EH6}. In the year of 2012, the authors in \cite{EH7} first proposed two practically realizable receiver architecture designs, in which EH and information detection could be operated in time switching (TS) or power splitting (PS) patterns. So far, these two receiver schemes have been widely adopted and used in various wireless systems \cite{EH_ONE_WAY}-\cite{EH_MULTI_COOPERA}. Considering that the two receiver designs are easy to be implemented in practical systems, we design the TSCMT and PSCMT protocols on the basis of TS and PS receiver architecture respectively.}.
\subsection{Protocol Description}

\begin{figure}
\centering
\includegraphics[width=0.45\textwidth]{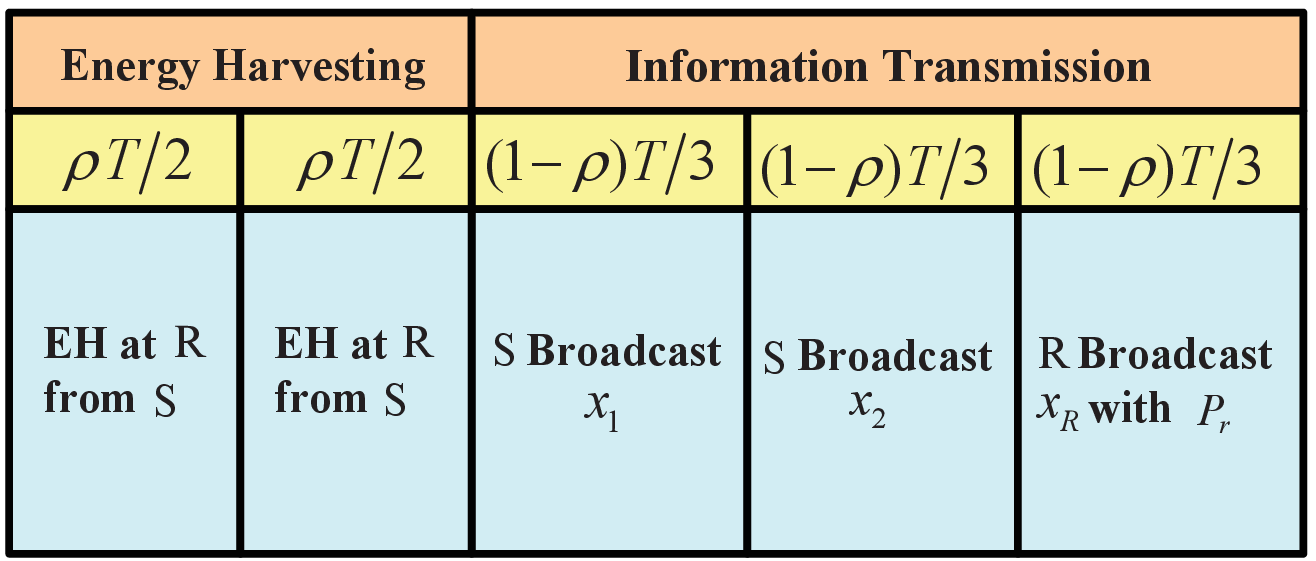}
\caption{Key parameters in the proposed TSCMT protocol.}
\label{TS_protocol}
\end{figure}

Fig.~\ref{TS_protocol} depicts the transmission process and key parameters in the proposed TSCMT protocol. For a time period $T$, let ${\rm{0}} \le \rho \le 1$ denote the time assignment factor, such that $\rho T$ part is assigned for ${\rm{R}}$ to harvest energy from ${\rm{S}}$, where it is equally divided into two time durations. Each $\rho T/2$ duration is assigned for  ${\rm{R}}$ to harvest energy from ${\rm{S}}$ during the period when ${\rm{S}}$ broadcasts $x_i$ ($i=1,2$) with power ${P_i}$. The remaining part $(1-\rho) T$ is used for the information transmission, which is equally divided into three parts. During the first two $(1-\rho) T/3$ durations, ${\rm{S}}$ broadcasts information $x_i$ with power ${P_i}$, both ${\rm{D_1}}$, ${\rm{D_2}}$ and ${\rm{R}}$ can receive the signal. In the third time duration of $(1-\rho) T/3$, ${\rm{R}}$ first combines the two signals it received, and then uses all the energy harvested from ${\rm{S}}$ to broadcast the combined signal $x_R$.
In the following subsection, we will analyze the system outage performance for the TSCMT protocol.

\subsection{Outage Probability Analysis for the TSCMT Protocol}\label{ts_out_analysis}
In this section, we will analyze the expression of the outage probability for the transmission from ${\rm{S}}$ to ${\rm{D_1}}$. Note that, the transmission from ${\rm{S}}$ to ${\rm{D_2}}$ has the similar outage performance due to the symmetry of the system.

As illustrated in Fig.~\ref{TS_protocol}, ${\rm{S}}$ broadcasts its information  $x_i$\footnote{Both $x_1$ and $x_2$  have unit average power} with power ${P_i}$, both ${\rm{D_1}}$, ${\rm{D_2}}$ and ${\rm{R}}$ can receive it. The received signals $y_{s,i}$ at ${\rm{D}}_i$, and the received signals $r_{s,j}$ at ${\rm{D}}_j$ ( $j=1,2$ and $j \ne i$) are given as follows, respectively:
\begin{equation}
{y_{s,i}} = \sqrt {{P_i}} {h_{s,i}}{x_i} + {n_{s,i}},\label{ys}
\end{equation}
\begin{equation}
{r_{s,j}} = \sqrt {{P_i}} {h_{s,j}}{x_i} + {v_{s,j}}.\label{rs}
\end{equation}
where $n_{s,i}$  and $v_{s,j}$  are the additive white Gaussian noise (AWGN) at ${\rm{D}}_i$ and ${\rm{D}}_j$ with $n_{s,i} \sim \mathcal{CN}(0,1)$ and $v_{s,j} \sim \mathcal{CN}(0,1)$. Note that,  $y_{s,i}$ is the desired signal of  ${\rm{D}}_i$, whereas  $r_{s,j}$ is the signal that can help ${\rm{D}}_j$  decode $x_i$  from the mixed signal transmitted by ${\rm{R}}$  in the third phase.

After the processing of the relay receiver, the sampled baseband signal at ${\rm{R}}$ is given as follows
\begin{equation}
{y_{r,i}} = \sqrt {{P_i}} {h_{s,r}}{x_i} + {n_{r,i}},  \label{eq_yr}
\end{equation}
where  $n_{r,i}$ is the AWGN at ${\rm{R}}$ with $n_{r,i} \sim \mathcal{CN}(0,1)$. The energy that ${\rm{R}}$ harvests from ${\rm{S}}$ is given as follows \cite{EH5}, \cite{EH_ONE_WAY}-\cite{EH_MULTI_COOPERA}
\begin{equation}
{E_i} = \eta {P_i}{\left| {{h_{s,r}}} \right|^2} \cdot \frac{{{\rho}T}}{2},\label{eq_ei}
\end{equation}
where ${\rm{0}} \le \rho  \le 1$ denotes the time assignment factor, and $0 < \eta  \le 1$  denotes the energy conversion efficiency.

After the first two transmissions from ${\rm{S}}$, the relay has harvested total $E_1+E_2$ energy from ${\rm{S}}$. So, the transmit power at ${\rm{R}}$ in the following phase is given by
\begin{equation}
{P_r} = \frac{{{E_1} + {E_2}}}{{(1 - \rho )T/3}} = \frac{{3\rho }}{{2(1 - \rho )}} \cdot \eta ({P_1} + {P_2}){\left| {{h_{s,r}}} \right|^2}.
\end{equation}

${\rm{R}}$ first combines the two signals  $y_{r,1}$ and $y_{r,2}$  as $x_R$ \cite{Relay_selection,xiong_globe2013_AF}, and uses $P_r$  to broadcast the combined signal $x_R$. Specifically, $x_R$ is given as follows
\begin{equation}
{x_R} = {\xi _1}{y_{r,1}} + {\xi _2}{y_{r,2}},\label{xr}
\end{equation}
where  $\xi _i$ ( $i=1,2$ ) denotes how  ${\rm{R}}$ combines  $y_{r,1}$ and $y_{r,2}$, and is selected as follows:
\begin{equation}
{\xi _i} = \sqrt {\frac{{{\theta _i}}}{{{P_i}{{\left| {{h_{s,r}}} \right|}^2} + 1}}}  \approx \sqrt {\frac{{{\theta _i}}}{{{P_i}{{\left| {{h_{s,r}}} \right|}^2}}}},\label{eq_appro}
\end{equation}
where $0 < {\theta _i} < 1$ ($i=1,2$ ), and $\theta _1+\theta _2=1$. The approximation in (7) is widely adopted in similar articles \cite{EH_TWO_WAY},\cite{Relay_selection}. Note that, $x_R$ always has unit power irrespective of $\theta _i$.

After combining the two signals  $y_{r,1}$ and $y_{r,2}$,  ${\rm{R}}$ broadcasts  $x_R$ with power $P_r$, and the signals received by $D_1$  and  $D_2$ are given as follows:
\begin{equation}
{y_{d,i}} = \sqrt {{P_r}} {h_{r,i}}{x_R} + {n_{d,i}},\label{eq_yd}
\end{equation}
where  $n_{d,i}$ is the AWGN at  ${\rm{D}}_i$ ( $i=1,2$) with $n_{d,i} \sim \mathcal{CN}(0,1)$. Because ${\rm{D}}_1$ ( ${\rm{D}}_2$) can decode   $x_2$ ( $x_1$) from (2), it can remove $x_2$ ( $x_1$ ) from $n_{d,1}$ ( $n_{d,2}$). Thus,  ${\rm{D}}_1$ can obtain the interference-free signal as follows:
\begin{eqnarray}
{\tilde y_{d,1}} &=& \sqrt {{P_r}} {h_{r,1}}{\xi _1}\sqrt {{P_1}} {h_{s,r}}{x_1} + \sqrt {{P_r}} {h_{r,1}}{\xi _2}{n_{r,2}} \nonumber\\ & & + \sqrt {{P_r}} {h_{r,1}}{\xi _1}{n_{r,1}} + {n_{d,1}}.
\end{eqnarray}

Then, submitting  $P_r$ in (5), and using the approximation in (7), the instantaneous SNR  $\gamma_1$ of the signal  $\tilde y_{d,1}$ is given as follows
\begin{equation}
{\gamma _1} = \frac{{\frac{{3\rho }}{{2(1 - \rho )}} \cdot \eta ({P_1} + {P_2}){\theta _1}{{\left| {{h_{s,r}}} \right|}^2}{{\left| {{h_{r,1}}} \right|}^2}}}{{\frac{{3\rho }}{{2(1 - \rho )}} \cdot \eta ({P_1} + {P_2})\left( {\frac{{{\theta _1}}}{{{P_1}}} + \frac{{{\theta _2}}}{{{P_2}}}} \right){{\left| {{h_{r,1}}} \right|}^2} + 1}}.
\end{equation}

By receiving two copies of $x_1$, ${\rm{D}}_1$  performs the maximal ratio combining (MRC). MRC is a method of diversity combining, in which the different copies of the same transmitted signal are added together to enhance the total received SNR at the destination \cite{xiong_TWC}, \cite{Relay_selection}. With the instantaneous SNR of the direct link from $\rm{S}$ to ${\rm{D}}_1$ which is denoted by ${\gamma _0} = {P_1}{\left| {{h_{s,1}}} \right|^2}$, we can obtain the mutual information of the transmission from $\rm{S}$ to ${\rm{D}}_1$ as follows
\begin{equation}
{I_1} = \frac{{2(1 - \rho )}}{3}\log (1 + {\gamma _0} + {\gamma _1}).
\end{equation}

As is known, the outage probability represents the probability which the target transmission rate is not supported due to the variations of channels. It is usually used to evaluate the performance over fading channels \cite{xiong_globe2013_AF}- \cite{xiong_TVT}. In our work,  an outage occur when the mutual information in (11) falls below the targeted rate $R_t$. Thus, the outage probability can be calculated as
\begin{eqnarray}
\lefteqn{{P_{\rm out}^{\rm (TS)}} = \Pr ({I_1} < {R_t})} \quad{\kern 1pt}{\kern 1pt}{\kern 1pt}{\kern 1pt}{\kern 1pt} \nonumber \\  & =& \Pr \left( {\frac{{2(1 - \rho )}}{3}\log (1 + {\gamma _0} + {\gamma _1}) < {R_t}} \right).\label{simu_ts}
\end{eqnarray}

\begin{figure}
\centering
\includegraphics[width=0.45\textwidth]{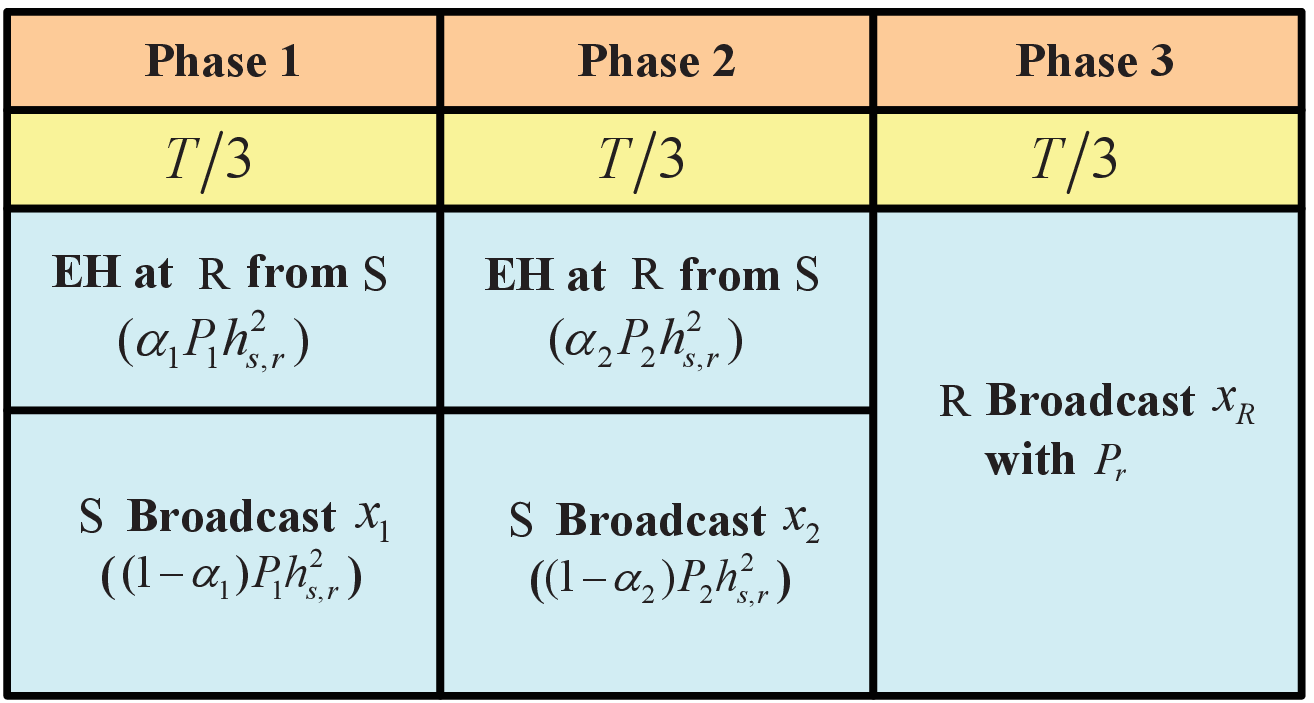}
\caption{Key parameters in the proposed PSCMT protocol.}
\label{PS_protocol}
\end{figure}

\textbf{Theorem 1}: Given a target transmission rate $R_t$, the outage probability of the TSCMT protocol for the multiuser cooperative transmission system with an EH relay is given as follows
\mathindent=0mm
\begin{flalign}
{P_{\rm out}^{\rm (TS)}} &= 1 - \exp \Big( - \frac{{{R_0}}}{{{P_1}{\Omega _{s,1}}}}\Big) + \sum\limits_{l = 1}^\infty  {\frac{{\exp \left( { - \frac{{{R_0}b}}{{a{\Omega _{s,r}}}}} \right){c^{l + 1}}}}{{{P_1}{\Omega _{s,1}}{{\left( {a{\Omega _{s,r}}{\Omega _{r,1}}} \right)}^{l + 1}}l!(l + 1)!}}} \nonumber \\  & \times \Bigg\{ \Bigg( {\ln \frac{c}{{a{\Omega _{s,r}}{\Omega _{r,1}}}} + 2\mathcal{C} - \sum\limits_{k = 1}^l {\frac{1}{k} - \sum\limits_{k = 1}^{l + 1} {\frac{1}{k}} } } \Bigg)\nonumber\\& \times{{\left( {\frac{1}{{{P_1}{\Omega _{s,1}}}} - \frac{b}{{a{\Omega _{s,r}}}}} \right)}^{ - l - 2}} \times \gamma \Big(l + 2,\frac{{{R_0}}}{{{P_1}{\Omega _{s,1}}}} - \frac{{b{R_0}}}{{a{\Omega _{s,r}}}}\Big)\nonumber\\ &+ {H_l}\Bigg\} +\frac{{\exp \left( { - \frac{{{R_0}}}{{{P_1}{\Omega _{s,1}}}}} \right)c}}{{a{\Omega _{s,r}}{\Omega _{r,1}}{P_1}{\Omega _{s,1}}}}\Bigg\{ \Bigg(\ln \frac{c}{{a{\Omega _{s,r}}{\Omega _{r,1}}}} + 2\mathcal{C}\Bigg)\nonumber\\ & \times {\Bigg(\frac{1}{{{P_1}{\Omega _{s,1}}}} - \frac{b}{{a{\Omega _{s,r}}}}\Bigg)^{ - 2}} \times \gamma \Big(2,\frac{{{R_0}}}{{{P_1}{\Omega _{s,1}}}} - \frac{{b{R_0}}}{{a{\Omega _{s,r}}}}\Big) + {H_0}\Bigg\} \nonumber\\&- \frac{{a{\Omega _{s,r}}}}{{a{\Omega _{s,r}} - b{P_1}{\Omega _{s,1}}}}\left( {\exp \Big( - \frac{{{R_0}}}{{{P_1}{\Omega _{s,1}}}}\Big) - \exp\Big( - \frac{{{R_0}b}}{{a{\Omega _{s,r}}}}\Big)} \right). \label{outage_annalysis}
\end{flalign}
\mathindent=7mm
where $\gamma \left( \cdot \right)$ denotes the incomplete gamma function, ${R_0} = {2^{\frac{{1.5{R_t}}}{{1 - \rho }}}} - 1$, $a = \frac{3}{2}\rho {(1 - \rho )^{ - 1}}\eta ({P_1} + {P_2}){\theta _1}$, $b = aP_2^{ - 1} + a{\theta _1}P_1^{ - 1}\theta _2^{ - 1}$, $c=1$, and $\mathcal{C}\approx 0.5772$ is Euler's constant. Here we have
\begin{eqnarray}
\mathcal{H}_{l} = \int_0^{{R_0}} {\exp ( - \frac{{\left( {a{\Omega _{s,r}} - b{P_1}{\Omega _{s,1}}} \right)\tau }}{{a{P_1}{\Omega _{s,1}}{\Omega _{s,r}}}})} {\tau ^{l + 1}}\ln \tau d\tau.
\end{eqnarray}
\textbf{Proof}: See Appendix A. \QED
\section{Power Splitting-based Cooperative Multiuser Transmission (PSCMT) Protocol}
\label{PSCMT}
In this section, by considering the power splitting receiver architecture, we shall first detail the proposed PSCMT protocol, and then analyze the system outage performance for it.
\subsection{Protocol Description}
Fig.~\ref{PS_protocol} illustrates the transmission process and key parameters in the proposed power splitting-based cooperative multiuser transmission (PSCMT) protocol. As shown in the figure, the transmission is accomplished through three phases, and each phase lasts for a time duration of $T/3$. During the $i$-th ($i=1,2$) phase, ${\rm{S}}$  broadcasts its information $x_i$ with power ${P_i}$, both ${\rm{D_1}}$, ${\rm{D_2}}$ and ${\rm{R}}$ can receive the signal. The part $(1 - {\alpha _i}){P_i}h_{s,r}^2$ is used for the information transmission from ${\rm{S}}$  to  ${\rm{R}}$, where $0 \le {\alpha _i} \le 1$ denotes the power splitting factor, and the other part $ {\alpha _1}{P_1}h_{s,r}^2$ is used for energy harvesting at ${\rm{R}}$. In the third phase, the relay first combines the two signals it received in the first two phases, and then uses the energy harvested from  ${\rm{S}}$ to broadcast the combined signal $x_R$ in a time duration of $T/3$.
\subsection{Outage Probability Analysis for the PSCMT Protocol}
As illustrated in Fig.~\ref{PS_protocol}, during the $i$-th ($i=1,2$) phase, ${\rm{S}}$  broadcasts its information $x_i$ with power ${P_i}$. The received signals  $y_{s,i}$ at ${\rm{D}}_i$, and the received signals $r_{s,j}$ at ${\rm{D}}_j$ ( $j=1,2$ and $j \ne i$) are given in (\ref{ys}) and (\ref{rs}), respectively.

At the end of the  $i$-th phase, after the processing of the relay receiver, the sampled baseband signal at ${\rm{R}}$ is given as follows
\begin{equation}
{y_{r,i}} = \sqrt {(1 - {\alpha _i}){P_i}} {h_{s,r}}{x_i} + {n_{r,i}},
\end{equation}
where $0 \le {\alpha _i} \le 1$ ($i=1,2$) denotes the power splitting factor, and $n_{r,i}$  is defined below (\ref{eq_yr}). The energy that ${\rm{R}}$  harvests from ${\rm{S}}$  is given by
\begin{equation}
{E_i} = \eta {P_i}{\alpha _i}{\left| {{h_{s,r}}} \right|^2} \cdot \frac{T}{3},
\end{equation}
where  $\eta$ is defined below (\ref{eq_ei}).

After the first two phases, the relay has harvested total $E_1+E_2$  energy from ${\rm{S}}$. So, the transmit power at ${\rm{R}}$  in the third phase is given by
\begin{equation}
{P_r} = \frac{{{E_1} + {E_2}}}{{T/3}} = \eta ({\alpha _1}{P_1} + {\alpha _2}{P_2}){\left| {{h_{s,r}}} \right|^2}.\label{eq_pr_nc}
\end{equation}

During the third phase, ${\rm{R}}$  first combines the two signals $x_1$  and $x_2$  as  $x_R$, which is given in (\ref{xr}), and uses the power $P_r$  to broadcast the combined signal $x_R$.

At the end of the third phase, the signals received by ${\rm{D}}_1$  and  ${\rm{D}}_2$  are given in (\ref{eq_yd}). Similar to the process in the TSCMT protocol described in subsection~\ref{ts_out_analysis}, after removing the interference signal $y_{d,2}$, ${\rm{D}}_1$ can obtain the interference-free signal as follows:
\begin{eqnarray}
{\tilde y_{d,1}} &=& \sqrt {{P_r}} {h_{r,1}}{\xi _1}\sqrt {(1 - {\alpha _1}){P_1}} {h_{s,r}}{x_1} + \sqrt {{P_r}} {h_{r,1}}{\xi _2}{n_{r,2}} \nonumber\\ & & + \sqrt {{P_r}} {h_{r,1}}{\xi _1}{n_{r,1}} + {n_{d,1}}.\label{eq_yd_nointerfrence_ps}
\end{eqnarray}

Submitting $P_r$  from (\ref{eq_pr_nc}) into (\ref{eq_yd_nointerfrence_ps}), and using the approximation in (\ref{eq_appro}), the instantaneous SNR $\gamma_1$  of the signal $\tilde y_{d,1}$  is given as follows
\begin{equation}
{\gamma _1} = \frac{{\eta ({\alpha _1}{P_1} + {\alpha _2}{P_2}){\theta _1}{{\left| {{h_{s,r}}} \right|}^2}{{\left| {{h_{r,1}}} \right|}^2}}}{{\eta ({\alpha _1}{P_1} + {\alpha _2}{P_2}){{\left| {{h_{r,1}}} \right|}^2}\left( {\frac{{{\theta _1}}}{{{P_1}(1 - {\alpha _1})}} + \frac{{{\theta _2}}}{{{P_2}(1 - {\alpha _2})}}} \right) + 1}}.
\end{equation}

\begin{figure}
\centering
\includegraphics[width=0.35\textwidth]{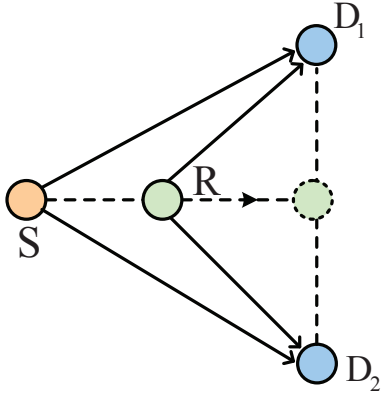}
\caption{Illustration of the relay position.}
\label{relay_position}
\end{figure}

Similar to the TSCMT protocol, ${\rm{D}}_1$  performs the maximal ratio combining (MRC). With the instantaneous SNR of the direct link from ${\rm{S}}$  to ${\rm{D}}_1$  which is denoted by ${\gamma _0} = {P_1}{\left| {{h_{s,1}}} \right|^2}$, the mutual information of the transmission from  ${\rm{S}}$  to ${\rm{D}}_1$ can be expressed as follows
\begin{equation}
{I_1} = \frac{2}{3}\log (1 + {\gamma _0} + {\gamma _1}),
\end{equation}
where the factor  $\frac{2}{3}$ in (19) is due to the fact that three phases are used to transmit two new signals. Thus, the outage probability can be calculated as
\begin{eqnarray}
\lefteqn{{P_{\rm out}^{\rm (PS)}} = \Pr ({I_1} < {R_t})} \quad{\kern 1pt}{\kern 1pt}{\kern 1pt}{\kern 1pt}{\kern 1pt} \nonumber \\  & =& \Pr \left( {\frac{2}{3}\log (1 + {\gamma _0} + {\gamma _1}) < {R_t}} \right).\label{simu_ps}
\end{eqnarray}
\textbf{Theorem 2}: Given a target transmission rate $R_t$, the outage probability of the PSCMT protocol for the multiuser cooperative transmission system can be analytically calculated using (\ref{outage_annalysis}), where\footnote{The detailed derivation of the outage probability for the PSCMT protocol is omitted here because it follows the same steps which is given below \ref{A}.} ${R_0} = {2^{1.5{R_t}}} - 1$, $a = \eta ({\alpha _1}{P_1} + {\alpha _2}{P_2}){\theta _1}$, $b = aP_2^{ - 1}{(1 - {\alpha _2})^{ - 1}} + a{\theta _1}\theta _2^{ - 1}P_1^{ - 1}{(1 - {\alpha _1})^{ - 1}}$, $c=1$.

 It is desirable to obtain the optimal values of $\rho$ and $\alpha$ which result in the lowest system outage probability ($\rho$ for the TSCMT protocol and $\alpha$ for the PSCMT protocol, respectively). But it is intractable to derive the closed-form expressions for the optimal $\rho$ and $\alpha$ due to the Bessel function and the integration involved in the explicit expression of $P_{\rm out}^{\rm (TS)}$ and $P_{\rm out}^{\rm (PS)}$, as shown in (\ref{outage_annalysis}). However, for given system configuration parameters, such as the source power and relay location, the optimization can be done offline by numerically evaluating optimal values of $\rho$ and $\alpha$.
\section{Numerical Results}\label{simulation}
In this section, numerical results are provided to verify our theoretical analysis on the system outage probability of the two proposed protocols for the multiuser transmission system. Moreover, the effects of the source transmit power and relay position on the system outage performance will be discussed, based on which, the optimal values of $\rho$ and $\alpha$ are numerically obtained.

Unless specifically stated, we set $R_t=1$bit/sec/Hz, $\eta=1$, $P_1=P_2=P_s$, and  $m=4$ (which corresponds to the urban areas and is widely adopted in literatures \cite{Relay_selection},\cite{Commun_book}). Both  $d_{s,1}$ and $d_{s,2}$  are normalized to 1. For simplicity, we set $\alpha_1=\alpha_2=\alpha$ . The distance between  ${\rm{D}}_1$  and  ${\rm{D}}_2$  is normalized to 1, and  ${\rm{R}}$  is placed on the height of the triangle composed of ${\rm{S}}$,  ${\rm{D}}_1$  and  ${\rm{D}}_2$, as illustrated in Fig.~\ref{relay_position}.

\begin{figure}
\centering
\includegraphics[width=0.48\textwidth]{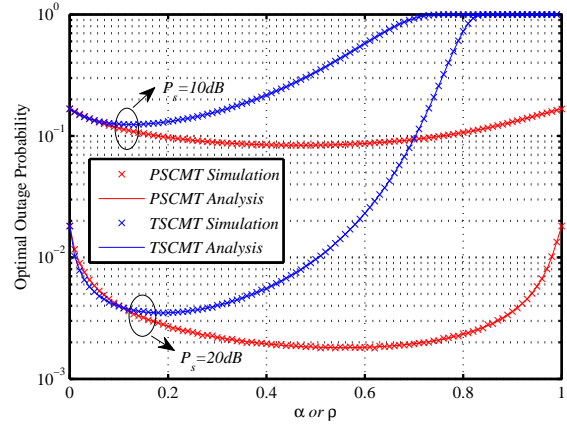}
\caption{Outage probability: numerical vs simulation. Other parameter: $d_{s,r}=0.5$.}
\label{fig_compare}
\end{figure}
\subsection{Verification of the Analytical Outage Probability}
In this subsection, simulation results are obtained through the Monte Carlo simulation using (\ref{simu_ts}) and (\ref{simu_ps}) to check the accuracy of our analytical expressions for the outage probability in Theorem 1 and Theorem 2. As shown in Fig.~\ref{fig_compare}, the simulation results closely match with the analytical results for all $\rho$ of the TSCMT protocol, and for all $\alpha$  of the PSCMT protocol, which verifies the analytical expressions for the outage probability of the two proposed protocols.

It can also be obtained that, for the same transmit power $P_s$  at ${\rm{S}}$, the PSCMT protocol outperforms the TSCMT protocol in terms of optimal outage probability.
\subsection{Effect of Source Power on Outage Probability}
\begin{figure*}[tb]
\centering
\subfigure[ ]{\label{fig_effectP_a} 
\includegraphics[width=0.48\textwidth]{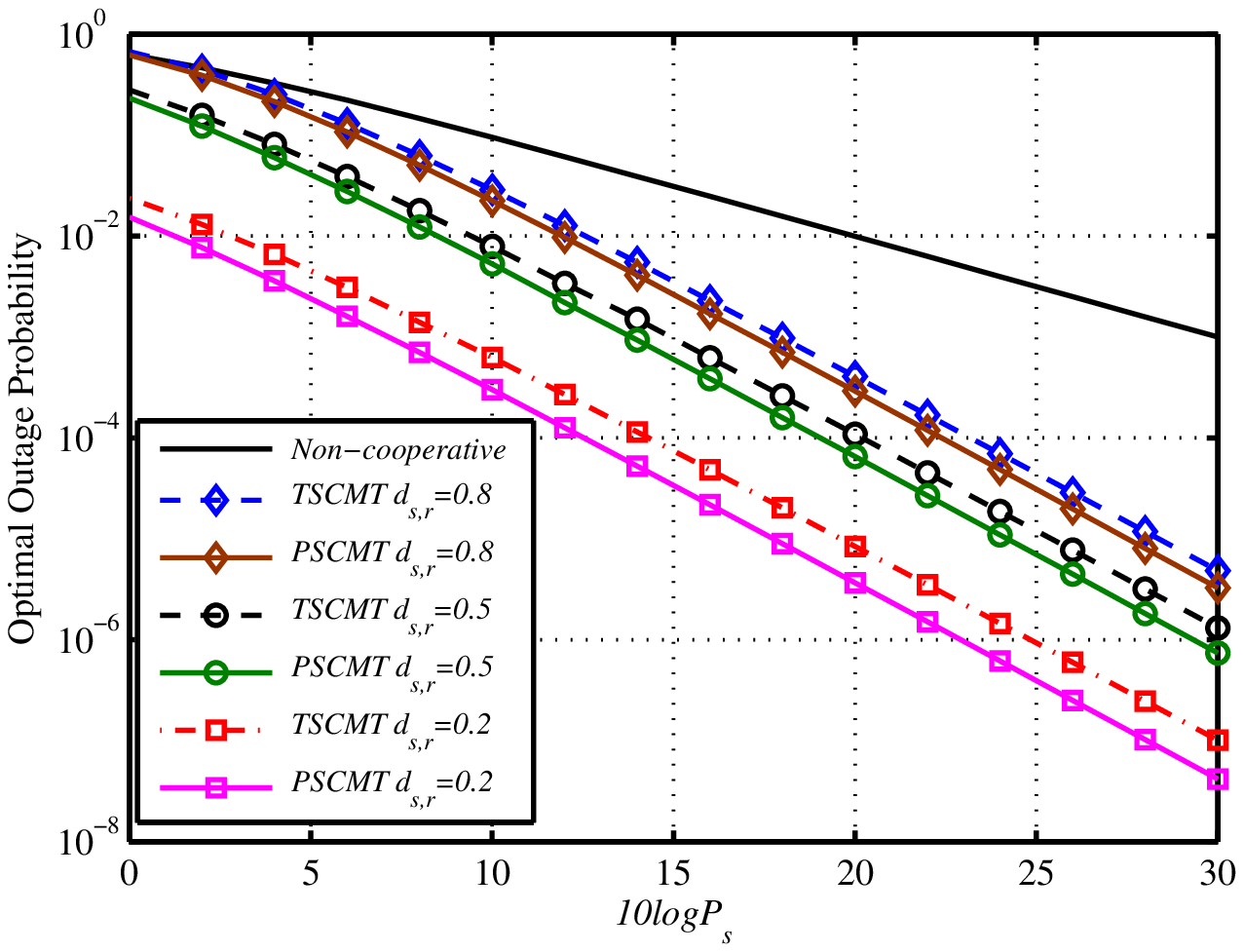}}
\subfigure[ ]{\label{fig_effectP_b} 
\includegraphics[width=0.48\textwidth]{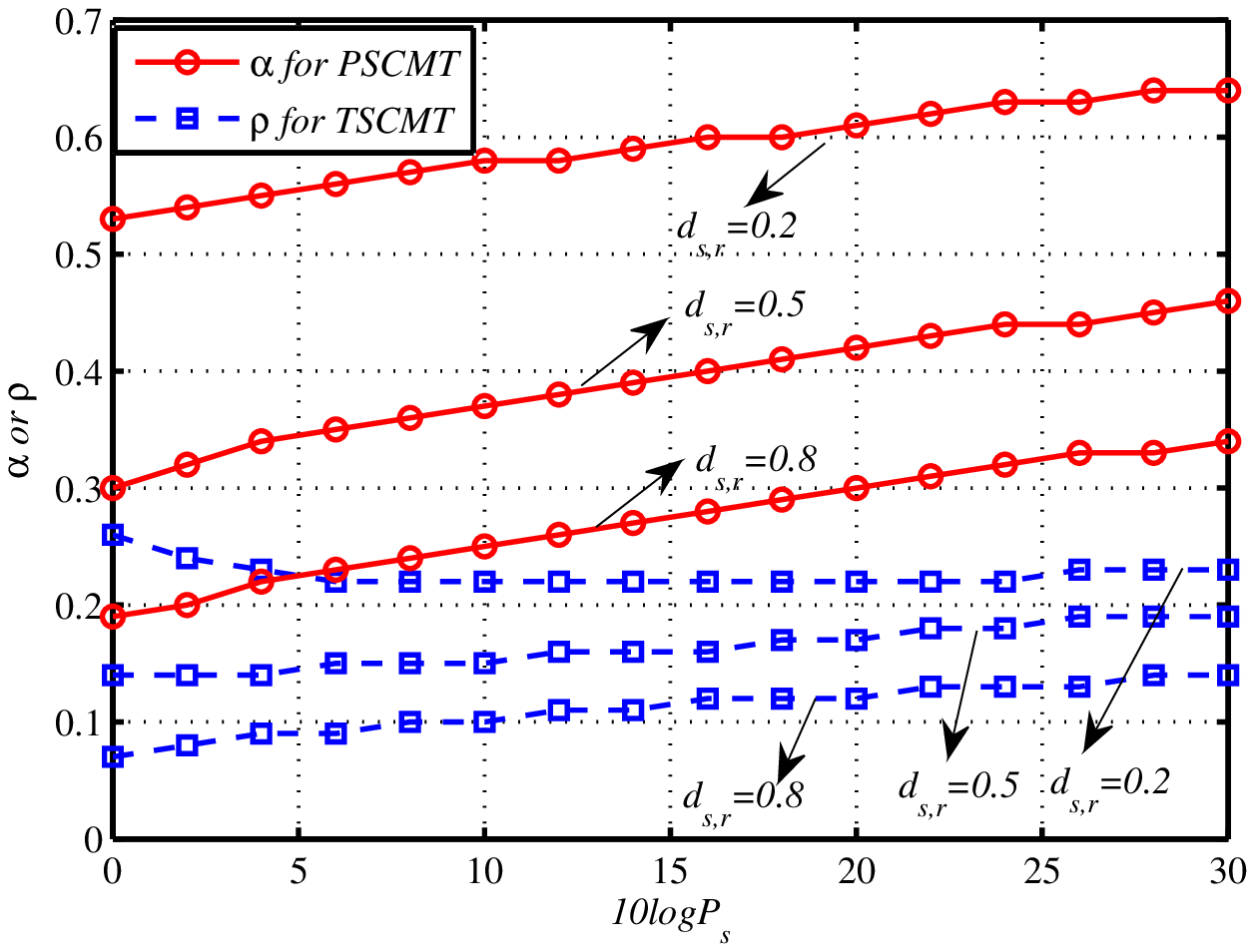}}

\caption{(a) Optimal system outage probability against $10{\log _{10}}{P_s}$ (b) optimal $\alpha$ or $\rho$  against $10{\log _{10}}{P_s}$.}
\label{fig_effectP} 
\end{figure*}

Fig.~\ref{fig_effectP_a} shows the optimal outage probability against $10{\log _{10}}{P_s}$ \footnote{Because the variances of the noise at all terminals are normalized to 1, $10{\log _{10}}{P_s}$ actually denotes the signal-to-noise ratio (SNR).} of the two proposed protocols. We also compare the outage probability of the traditional non-cooperative scheme to the two proposed protocols. It can be seen that, the two proposed protocols are better than the traditional non-cooperative scheme to obtain lower outage probability. Moreover, in higher  $10{\log _{10}}{P_s}$ region, as $10{\log _{10}}{P_s}$  increases, the outage probability of the TSCMT protocol and the PSCMT protocol decrease faster than the non-cooperative scheme. Besides, for the same $10{\log _{10}}{P_s}$, the PSCMT protocol outperforms the TSCMT protocol in terms of system outage probability in the whole $10{\log _{10}}{P_s}$  region.

Fig.~\ref{fig_effectP_b} shows the optimal  $\alpha$ (for the PSCMT protocol) and  $\rho$ (for the TSCMT protocol) versus $10{\log _{10}}{P_s}$. As shown in Fig.~\ref{fig_effectP_b}, the optimal  $\alpha$ increases as $10{\log _{10}}{P_s}$  increases. This is due to the fact that, when the SNR is higher, the relay need less power to process information, thus more energy is left to be harvested.
\subsection{Effect of Relay Location on Outage Probability}
\begin{figure*}[tb]
\centering
\subfigure[ ]{\label{fig_effectD_a} 
\includegraphics[width=0.48\textwidth]{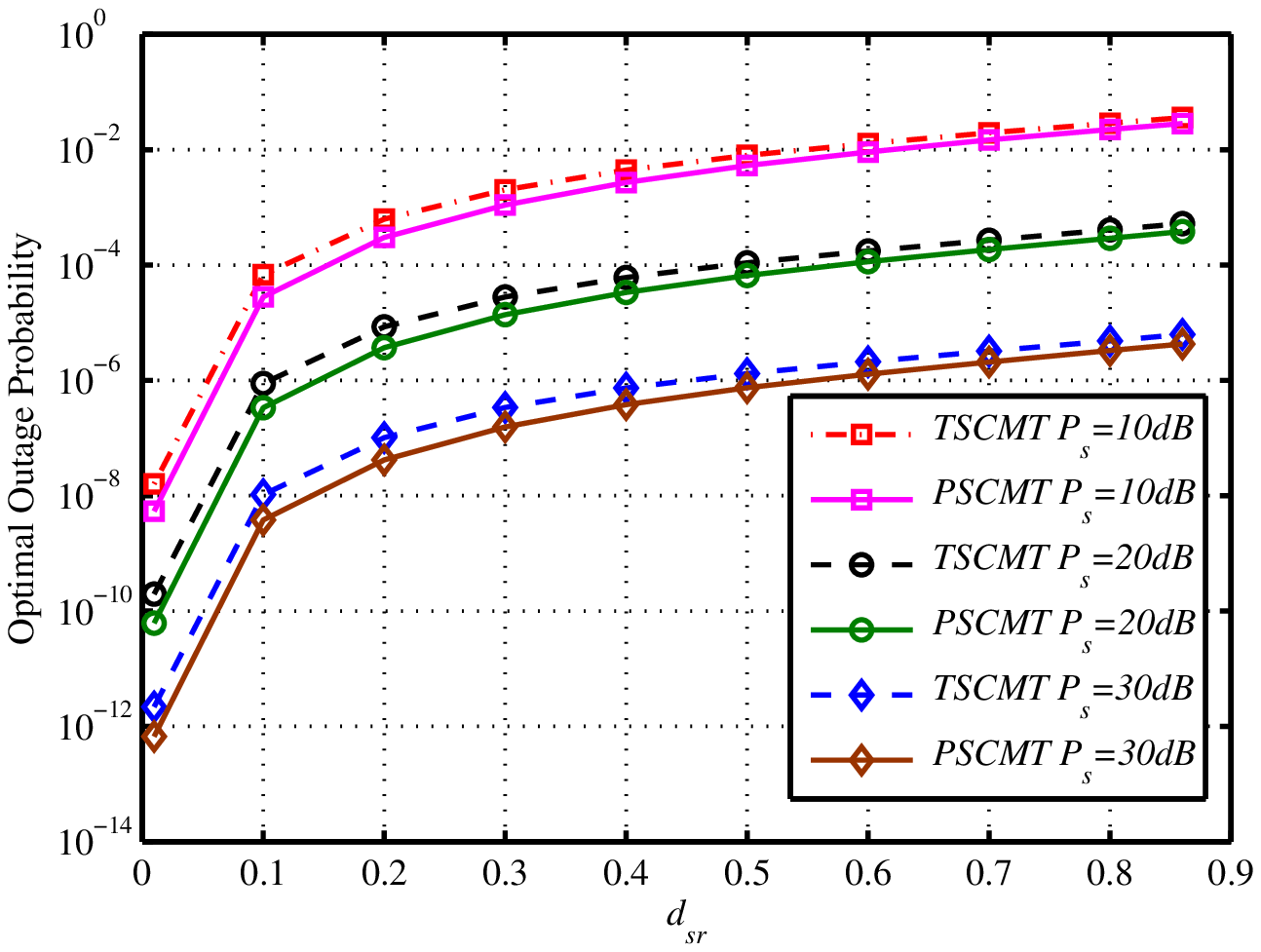}}
\subfigure[ ]{\label{fig_effectD_b} 
\includegraphics[width=0.48\textwidth]{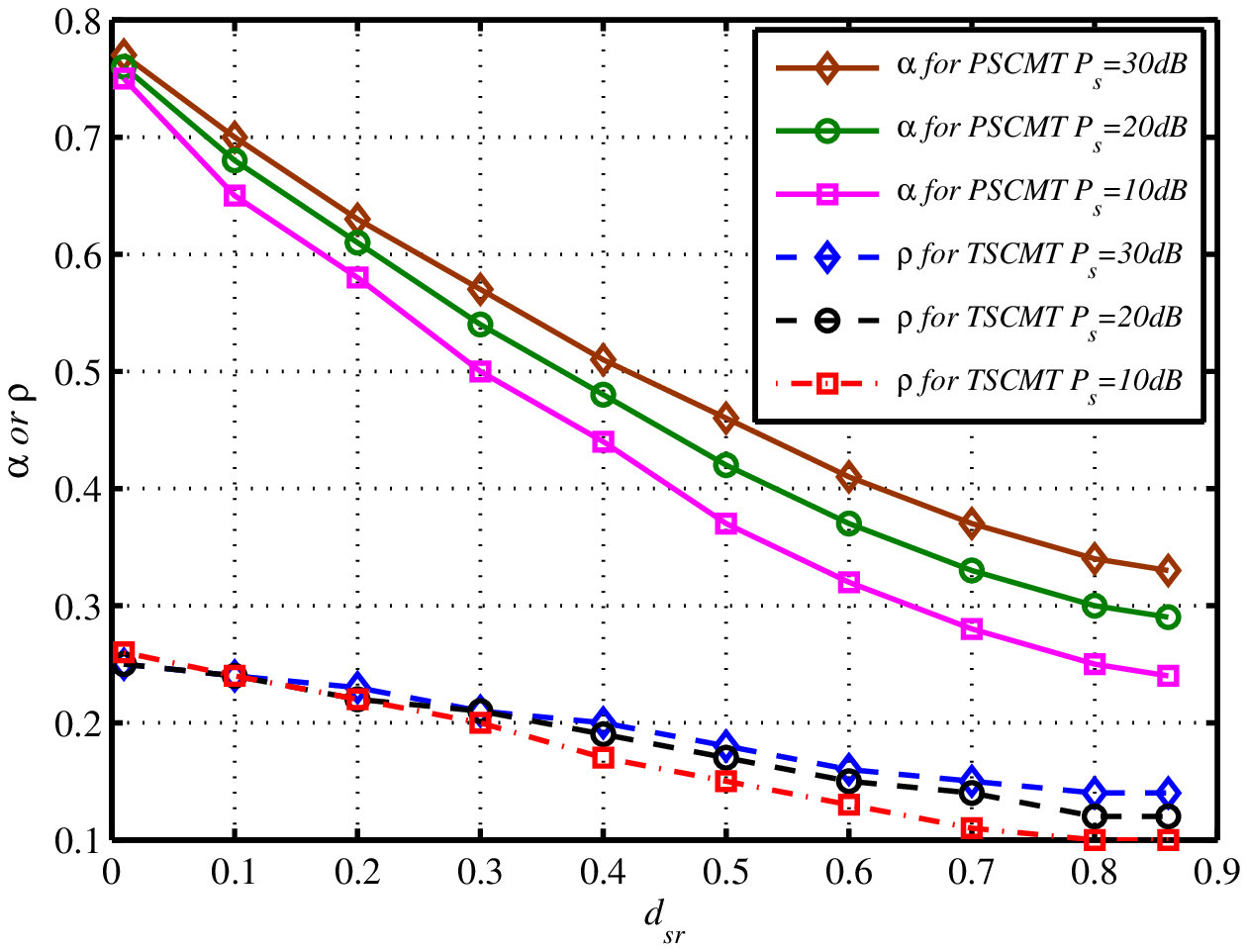}}

\caption{(a) Optimal system outage probability against $d_{s,r}$ (b) optimal $\alpha$ or $\rho$  against $d_{s,r}$.}
\label{fig_effectD} 
\end{figure*}

Fig.~\ref{fig_effectD_a} shows the optimal outage probability versus $d_{s,r}$. As shown in Fig.~\ref{fig_effectD_a}, the optimal outage probability increases as $d_{s,r}$  increases, which indicates that to obtain lower outage probability, it is better to choose the relay near the source. For the relay with a particular $d_{s,r}$, the PSCMT protocol can always achieve lower system outage probability than the TSCMT protocol.

Fig.~\ref{fig_effectD_b} shows the optimal  $\alpha$ (for the PSCMT protocol) and  $\rho$ (for the TSCMT protocol) versus $d_{s,r}$. It can be observed that, the optimal  $\alpha$ and  $\rho$ decrease as  $d_{s,r}$ increases. This is due to the fact that, when the relay is far away from the source, it needs more power to correctly receive the information form the source, which makes less energy left to be harvested.
\section{Conclusion}\label{conclusion}
In this paper, we investigated the multiuser transmission network with an EH cooperative relay. By adopting the time switching and the power splitting relay receiver architectures, we first proposed two cooperative protocols for the multiuser transmission system: the TSCMT protocol and the PSCMT protocol. Then, for each proposed protocol, we derived the explicit expression for the system outage probability. Moreover, we also discussed the effects of various system parameters, such as the source's transmit power and the relay location on the system outage performance. Numerical results showed that our proposed protocols achieve lower system outage probability without consuming additional energy compared with traditional non-cooperative scheme. Besides, for the same transmit power at the source, the PSCMT protocol is superior to the TSCMT protocol to obtain lower system outage probability.

\section{Conflict of Interests}
The authors declare that there is no conflict of interests regarding the publication of this paper.
\section{Acknowledgments}
This work was supported by the Fundamental Research Funds for the Central Universities (no. 2014JBM024).
\appendix
\section{}\label{A}
This appendix derives the $P_{\rm out}^{\rm (TS)}$ in (\ref{outage_annalysis}) for the TSCMT protocol.

By denoting ${\left| {{h_{r,1}}} \right|^2}$ and  ${\left| {{h_{s,r}}} \right|^2}$ as $X$  and $Y$, we define the variable $Z = \frac{{aXY}}{{bX + c}} = {\gamma _1}$. We can see that $Z$ is a combination of two independent random variables. Using the basic knowledge of the probability theory, we can obtain the cumulative density function (CDF) ${F_Z}(z)$  of  $Z$ which is given by
\mathindent=0mm
\begin{eqnarray}
{F_Z}(z) = 1 - \frac{{\exp \left( { - \frac{{zb}}{{a{\Omega _{s,r}}}}} \right)}}{{{\Omega _{r,1}}}}\sqrt {\frac{{4zc{\Omega _{r,1}}}}{{a{\Omega _{s,r}}}}} {K_1}\left( {\sqrt {\frac{{4zc}}{{a{\Omega _{r,1}}{\Omega _{s,r}}}}} } \right), \label{Fz}
\end{eqnarray}
\mathindent=7mm
where ${K_1}\left(  \cdot  \right)$ denotes the first-order modified Bessel function of the second kind \cite{Integral}.

Since $\gamma_0$ is an exponentially distributed random variable with mean ${P_1}{\Omega _{{\rm{S}},1}}$, $P_{\rm out}^{\rm (TS)}$ can be rewritten as follows
\begin{eqnarray}
\lefteqn{{P_{\rm out}^{\rm (TS)}} = \Pr ({\gamma _0} + {\gamma _1} < {R_0})} \quad{\kern 1pt}{\kern 1pt}{\kern 1pt}{\kern 1pt}{\kern 1pt} \nonumber \\  & =& \int_0^{{R_0}} {\Pr [Z < {R_0} - \tau ]} {\kern 1pt} {\kern 1pt} {f_{{\gamma _0}}}(\tau )d\tau\nonumber \\  & =& \int_0^{{R_0}} {{F_Z}({R_0} - \tau )} \frac{1}{{{P_1}{\Omega _{{\rm{S}},1}}}}\exp ( - \frac{\tau }{{{P_1}{\Omega _{{\rm{S}},1}}}})d\tau. \label{integral}
\end{eqnarray}

The integral in (\ref{integral}) can't be directly calculated due to the ${K_1}\left(  \cdot  \right)$ in (\ref{Fz}). By applying the series expansion of ${K_1}\left(  \cdot  \right)$ which is given in (\ref{K_1}), we can rewritten $P_{\rm out}^{\rm (TS)}$  in (\ref{series}) as follows

\begin{flalign}
{K_1}\left( x \right) &= \sum\limits_{l = 0}^\infty  {\frac{{{{\left( {\frac{x}{2}} \right)}^{2l + 1}}}}{{l!(l + 1)!}}\left( {\ln \frac{x}{2} + \mathcal{C}} \right)} \nonumber \\ &  - \frac{1}{2}\sum\limits_{l = 1}^\infty  {\frac{{{{\left( {\frac{x}{2}} \right)}^{2l + 1}}}}{{l!(l + 1)!}}} \left( {\sum\limits_{k = 1}^l {\frac{1}{k} + \sum\limits_{k = 1}^{l + 1} {\frac{1}{k}} } } \right) + \frac{1}{x},\label{K_1}
\end{flalign}

\mathindent=0mm
\begin{flalign}
{P_{\rm out}^{\rm (TS)}} &= \frac{1}{{P_1}{\Omega _{s,1}}}\Bigg\{ \int_0^{{R_0}} {\exp ( - \frac{\tau }{{{P_1}{\Omega _{s,1}}}})} d\tau \nonumber \\&- \int_0^{{R_0}} \frac{\exp \left( - \frac{({R_0} - \tau )b}{a{\Omega _{s,r}}}\right)}{\Omega _{r,1}} \sum\limits_{l = 1}^\infty  {\frac{{{({R_0} - \tau )}^{l + 1}}{\Omega _{r,1}}{c^{l + 1}}}{{{\left( {a{\Omega _{s,r}}{\Omega _{r,1}}} \right)}^{l + 1}}l!(l + 1)!}} \nonumber \\&\times\Bigg( {\ln \frac{{({R_0} - \tau )c}}{{a{\Omega _{s,r}}{\Omega _{r,1}}}} + 2\mathcal{C} - \sum\limits_{k = 1}^l {\frac{1}{k} - \sum\limits_{k = 1}^{l + 1} {\frac{1}{k}} } } \Bigg)\exp ( - \frac{\tau }{{{P_1}{\Omega _{s,1}}}}) d\tau \nonumber \\&-\int_0^{{R_0}} \frac{{\exp \left( { - \frac{{({R_0} - \tau )b}}{{a{\Omega _{s,r}}}}} \right)({R_0} - \tau )}}{{a{\Omega _{s,r}}{\Omega _{r,1}}{P_1}{\Omega _{s,1}}}}\left( {\ln \frac{{({R_0} - \tau )c}}{{a{\Omega _{s,r}}{\Omega _{r,1}}}} + 2\mathcal{C}} \right)\nonumber \\&\times \exp ( - \frac{\tau }{{{P_1}{\Omega _{s,1}}}}) d\tau  - \int_0^{{R_0}} {\exp ( \frac{{( \tau- {R_0})b}}{{a{\Omega _{s,r}}}})\exp ( - \frac{\tau }{{{P_1}{\Omega _{s,1}}}})} d\tau \Bigg\}.\label{series}
\end{flalign}
\mathindent=7mm

By denoting the four integral items in the right hand side of the above equation as $Q_1$, $Q_2$, $Q_3$ and $Q_4$ respectively, we can obtain that
\begin{eqnarray}
{Q_1} = 1 - \exp \bigg( - \frac{{{R_0}}}{{{P_1}{\Omega _{s,1}}}}\bigg),\label{Q1}
\end{eqnarray}

\mathindent=0mm
\begin{flalign}
{Q_2} &=  - \sum\limits_{l = 1}^\infty  {\frac{{\exp \left( { - \frac{{{R_0}b}}{{a{\Omega _{s,r}}}}} \right){c^{l + 1}}}}{{{P_1}{\Omega _{s,1}}{{\left( {a{\Omega _{s,r}}{\Omega _{r,1}}} \right)}^{l + 1}}l!(l + 1)!}}} \nonumber \\ & \times  \int_0^{{R_0}} {\exp \bigg( - \frac{{\left( {a{\Omega _{s,r}} - b{P_1}{\Omega _{s,1}}} \right)\tau }}{{a{P_1}{\Omega _{s,1}}{\Omega _{s,r}}}}\bigg)} {\left( {{R_0} - \tau } \right)^{l + 1}} \nonumber \\ & \times \left\{ {\left( {\ln \frac{c}{{a{\Omega _{s,r}}{\Omega _{r,1}}}} + 2\mathcal{C} - \sum\limits_{k = 1}^l {\frac{1}{k} - \sum\limits_{k = 1}^{l + 1} {\frac{1}{k}} } } \right) + \ln ({R_0} - \tau )} \right\}d\tau \nonumber \\ &= \sum\limits_{l = 1}^\infty  {\frac{{\exp \left( { - \frac{{{R_0}b}}{{a{\Omega _{s,r}}}}} \right){c^{l + 1}}}}{{{P_1}{\Omega _{s,1}}{{\left( {a{\Omega _{s,r}}{\Omega _{r,1}}} \right)}^{l + 1}}l!(l + 1)!}}} \nonumber \\ &\times \Bigg\{ \Bigg( {\ln \frac{c}{{a{\Omega _{s,r}}{\Omega _{r,1}}}} + 2\mathcal{C} - \sum\limits_{k = 1}^l {\frac{1}{k} - \sum\limits_{k = 1}^{l + 1} {\frac{1}{k}} } } \Bigg)\nonumber \\ &\times {{\Bigg( {\frac{1}{{{P_1}{\Omega _{s,1}}}} - \frac{b}{{a{\Omega _{s,r}}}}} \Bigg)}^{ - l - 2}} \gamma \bigg(l + 2,\frac{{{R_0}}}{{{P_1}{\Omega _{s,1}}}} - \frac{{b{R_0}}}{{a{\Omega _{s,r}}}}\bigg)\nonumber \\ & + \int_0^{{R_0}} {\exp \Bigg( \frac{{\left( {b{P_1}{\Omega _{s,1}} -a{\Omega _{s,r}}} \right)\tau }}{{a{P_1}{\Omega _{s,1}}{\Omega _{s,r}}}}\Bigg)} {\tau ^{l + 1}}\ln \tau d\tau  \Bigg\},\label{Q2}
\end{flalign}
\mathindent=7mm
where $\gamma\left(  \cdot  \right)$  denotes the incomplete gamma function. $Q_3$  can be rewritten as

\mathindent=0mm
\begin{flalign}
{Q_3} &= \frac{{\exp \left( { - \frac{{{R_0}}}{{{P_1}{\Omega _{s,1}}}}} \right)c}}{{a{\Omega _{s,r}}{\Omega _{r,1}}{P_1}{\Omega _{s,1}}}}\Bigg\{\Bigg(\ln \frac{c}{{a{\Omega _{s,r}}{\Omega _{r,1}}}} + 2\mathcal{C}\Bigg){\Bigg(\frac{1}{{{P_1}{\Omega _{s,1}}}} - \frac{b}{{a{\Omega _{s,r}}}}\Bigg)^{ - 2}} \nonumber \\& \times \gamma \bigg(2,\frac{{{R_0}}}{{{P_1}{\Omega _{s,1}}}} - \frac{{b{R_0}}}{{a{\Omega _{s,r}}}}\bigg) + \int_0^{{R_0}} {\exp \bigg( \frac{{\left( {b{P_1}{\Omega _{s,1}} -a{\Omega _{s,r}}} \right)\tau }}{{a{P_1}{\Omega _{s,1}}{\Omega _{s,r}}}}\Bigg)}  \nonumber \\& \times\tau \ln \tau d\tau \Bigg\},\label{Q3}
\end{flalign}
\mathindent=7mm

\begin{flalign}
{Q_4} = & - \frac{{a{\Omega _{s,r}}}}{{a{\Omega _{s,r}} - b{P_1}{\Omega _{s,1}}}} \nonumber \\ & \times \left( {\exp \bigg( \frac{{{R_0}}}{{{P_1}{\Omega _{s,1}}}}-\frac{{2{R_0}b}}{{a{\Omega _{s,r}}}}\bigg) - \exp \bigg( - \frac{{{R_0}b}}{{a{\Omega _{s,r}}}}\bigg)} \right).\label{Q4}
\end{flalign}

Substituting (\ref{Q1}), (\ref{Q2}), (\ref{Q3}) and (\ref{Q4}) into (\ref{series}), one can obtain the final expression of $P_{\rm out}^{\rm (TS)}$, which is given in (\ref{outage_annalysis}). This ends the proof for Theorem 1. \QED

\end{document}